\documentstyle{mn}

\newif\ifAMStwofonts

\ifoldfss
  \ifCUPmtlplainloaded \else
    \NewTextAlphabet{textbfit} {cmbxti10} {}
    \NewTextAlphabet{textbfss} {cmssbx10} {}
    \NewMathAlphabet{mathbfit} {cmbxti10} {} 
    \NewMathAlphabet{mathbfss} {cmssbx10} {} 
  \fi
  \ifAMStwofonts
    \ifCUPmtlplainloaded \else
      \NewSymbolFont{upmath} {eurm10}
      \NewSymbolFont{AMSa} {msam10}
      \NewMathSymbol{\upi}     {0}{upmath}{19}
      \NewMathSymbol{\umu}     {0}{upmath}{16}
      \NewMathSymbol{\upartial}{0}{upmath}{40}
      \NewMathSymbol{\leqslant}{3}{AMSa}{36}
      \NewMathSymbol{\geqslant}{3}{AMSa}{3E}

      \let\leq=\leqslant 
      \let\geq=\geqslant 
    \fi
  \fi
\fi 

\ifnfssone
  \newmathalphabet{\mathit}
  \addtoversion{normal}{\mathit}{cmr}{m}{it}
  \addtoversion{bold}{\mathit}{cmr}{bx}{it}
  \newmathalphabet{\mathbfit} 
  \addtoversion{normal}{\mathbfit}{cmr}{bx}{it}
  \addtoversion{bold}{\mathbfit}{cmr}{bx}{it}
  \newmathalphabet{\mathbfss} 
  \addtoversion{normal}{\mathbfss}{cmss}{bx}{n}
  \addtoversion{bold}{\mathbfss}{cmss}{bx}{n}
  \ifAMStwofonts
    \ifCUPmtlplainloaded \else
      \UseAMStwoboldmath
      \makeatletter
      \new@mathgroup\upmath@group
      \define@mathgroup\mv@normal\upmath@group{eur}{m}{n}
      \define@mathgroup\mv@bold\upmath@group{eur}{b}{n}
      \edef\UPM{\hexnumber\upmath@group}
      \new@mathgroup\amsa@group
      \define@mathgroup\mv@normal\amsa@group{msa}{m}{n}
      \define@mathgroup\mv@bold\amsa@group{msa}{m}{n}
      \edef\AMSa{\hexnumber\amsa@group}
      \makeatother
      \mathchardef\upi="0\UPM19
      \mathchardef\umu="0\UPM16
      \mathchardef\upartial="0\UPM40
      \mathchardef\leqslant="3\AMSa36
      \mathchardef\geqslant="3\AMSa3E

      \let\leq=\leqslant 
      \let\geq=\geqslant 
    \fi
  \fi
\fi 

\ifnfsstwo
  \DeclareMathAlphabet{\mathbfit}{OT1}{cmr}{bx}{it}
  \SetMathAlphabet\mathbfit{bold}{OT1}{cmr}{bx}{it}
  \DeclareMathAlphabet{\mathbfss}{OT1}{cmss}{bx}{n}
  \SetMathAlphabet\mathbfss{bold}{OT1}{cmss}{bx}{n}
  \ifAMStwofonts
    \ifCUPmtlplainloaded \else
      \DeclareSymbolFont{UPM}{U}{eur}{m}{n}
      \SetSymbolFont{UPM}{bold}{U}{eur}{b}{n}
      \DeclareSymbolFont{AMSa}{U}{msa}{m}{n}
      \DeclareMathSymbol{\upi}{0}{UPM}{"19}
      \DeclareMathSymbol{\umu}{0}{UPM}{"16}
      \DeclareMathSymbol{\upartial}{0}{UPM}{"40}
      \DeclareMathSymbol{\leqslant}{3}{AMSa}{"36}
      \DeclareMathSymbol{\geqslant}{3}{AMSa}{"3E}

      \let\leq=\leqslant 
      \let\geq=\geqslant 
    \fi
  \fi
\fi 

\ifCUPmtlplainloaded \else
  \ifAMStwofonts \else 
    \def\upi{\pi}
    \def\umu{\mu}
    \def\upartial{\partial}
  \fi
\fi

\title{The effects of seeing on S\`ersic profiles}
\author[I. Trujillo et al.]
       {I. Trujillo$^{1}$, J. A. L. Aguerri$^{2}$, J. Cepa$^{1}$ and  C. M. Guti\'errez$^{1}$\\
        $^{1}$ Instituto de Astrof\'{\i}sica de Canarias,  E-38205 La Laguna, Tenerife, Spain\\
	$^{2}$ Astronomisches Institut der Universit\"{a}t Basel, Venusstrasse 7, CH-4102 Binningen, Switzerland}
\date{Accepted 0000 December 00.
      Received 0000 December 00;
      in original form 0000 October 00}

\pagerange{\pageref{firstpage}--\pageref{lastpage}}
\pubyear{1994}

\begin{document}

\maketitle

\label{firstpage}

\begin{abstract}
The effects of seeing on the parameters of the S\`ersic profile 
 are studied 
 in an analytical form using a Gaussian point spread function. The 
 surface brightness of S\`ersic profiles is proportional (in magnitudes) 
 to $r^{1/n}$. The parameter $n$ serves to classify the type of profile 
 and is related to the central luminosity concentration. It is the
parameter most 
 affected  by seeing; furthermore, the value of $n$ that 
 can be measured is always smaller than the real one. It is shown that 
 the luminosity density of the S\`ersic profile with $n$ less than 0.5 has 
 a central depression, which is physically unlikely. Also, the intrinsic 
 ellipticity of the sources has been taken into account and we show 
 that the parameters are dependent
when the effects of seeing are non-negligible. Finally, a prescription for correcting raw effective radii, central
 intensities and $n$ parameters is given.
\end{abstract}

\begin{keywords}
atmospheric effects -- methods: data analysis -- galaxies: distances and redshifts -- galaxies: photometry.
\end{keywords}

\section{Introduction}

Galaxies exhibit many different morphologies 
and therefore dynamical properties.  They also vary enormously in  size 
and luminosity.  Galaxies can consist of one or more dynamical 
substructures. The classification methods are mainly based on the morphologies 
of these substructures.   The most widely used classification scheme is that
 introduced by Hubble (1936), which is based on the ratio of 
 the spheroidal bulge and disc luminosities. The galactic morphological types 
 vary from ellipticals (with only  spheroidal components) to late-type spirals,
  with small spheroids and prominent disc components.  Irregular galaxies, 
  which are later than spirals in the Hubble sequence, 
are characterized by the absence of
   symmetry.   In recent decades much work has been done to improve 
   the method for determining the principal components of galaxies (e.g.  
   Prieto et al. 2000 and references therein).

It is usually assumed that the light of a galaxy follows the mass distribution.
 The mass distribution can then be inferred by modelling the light distribution.
  The light of a galaxy is usually modelled by fitting the surface brightness 
  profile of each structural component with certain analytical curves. These 
  laws have certain free parameters that must be determined during the fitting 
  process.  From de Vaucouleurs (1948), it is well known that the surface 
  brightness profile (in magnitudes) of elliptical galaxies is proportional 
  to $r^{1/4}$ (where $r$ is the radial distance to the centre of the galaxy).
    This law was also applied to bulges of spiral galaxies which have similar
     shapes, colours and kinematics to those of
 ellipticals.  It has recently been
     discovered that not all bulges follow an $r^{1/4}$ profile (Caon,
Capaccioli \& D'Onofrio
      1993; Andredakis, Peletier \& Balcells 1995). A better analytical 
form is the S\`ersic 
      profile (S\`ersic 1968) which has surface brightness proportional to 
      $r^{1/n}$ and generalizes the
$r^{1/4}$ law. The parameter $n$ is one of the three free parameters and 
it defines the type of the profile. When $n=1$ the surface brightness profile
 is exponential. Increasing values of $n$ give more centrally concentrated 
 luminosity profiles. Andredakis et al. (1995) found a correlation between 
 the value of $n$ and the morphological type of the galaxies, in the sense
  that early types show larger values of $n$ than do late-type galaxies. 
  Exponential profiles have been used extensively in order to fit the
   surface brightness profiles of discs of spiral galaxies. S\`ersic profiles 
   have been also used in  other types of galaxies, for example, Davies et
    al. (1988) propose that dwarf ellipticals are well fitted with a S\`ersic 
    law with $n\leq 2$.

Ground-based astronomical images are always affected by atmospheric blurring.
  Many papers have been written on the subject and much effort has been 
  invested in the construction of new optics to minimize its effects.  
  Seeing scatters light from the inner, centrally concentrated core to 
  the outer, more diffuse regions of galaxies, producing a mean surface 
  brightness lower than the true values and larger effective radii.  
  These effects can change the results of the photometric parameters 
  obtained from the fits of the observed surface brightness profiles;
   moreover, the dynamical properties inferred from these parameters will also
    be wrong.  Although seeing affects all points in a galaxy, its 
    effects are more important in the central regions. Seeing effects
     were extensively studied in the case of elliptical galaxies with 
     $r^{1/4}$ profiles (Franx, Illingwort \& Heckman 1989; Saglia et al. 
1993).  These authors 
     showed that the effects of seeing on the photometric properties of 
     elliptical galaxies can extend as far as 5 seeing discs.

Saglia et al. (1993) also showed that seeing effects become important 
for distant ellipticals ($cz > 8000$ km s$^{-1}$).  They found 
that uncorrected fundamental plane distances are systematically too 
high if  seeing effects are not taken into account.  This is due 
to the small angular size of the objects at high distances: for a flat
 universe with $H_{0}=75$ km s$^{-1}$ Mpc$^{-1}$, an object of 30 kpc has
  an angular size of 15$''$ at $z=0.1$ and 3$''$ at $z=0.5$.  This means 
  that a typical seeing of 1$''$ is equivalent to $1/15$ of the size of the
   object at $z=0.1$ but  is $1/3$ of the object at $z=0.5$.  The seeing 
will therefore produce important effects on these objects
 at large radii. 
   Thus, the study of the effects of seeing
 on  surface brightness profiles must 
   be taken into account when the photometric parameters of these galaxies
    are obtained from the decomposition of their surface brightness profiles.
      The new generation of ground-based telescopes and the study of galaxies
       at higher redshifts make these kinds of studies very important. 

In this paper we present an analytical treatment of  seeing effects on
 S\`ersic profiles taking into account the ellipticity of the objects. The 
 mathematical treatment of  seeing will be given in Section 2. In Section 
 3 we present the effect of the seeing on the photometric parameters. An easy
 prescription for correcting the parameters measured from the raw profiles is given
 in Section 4. 
 The analysis of these results are given in Section 5.

\section[]{Mathematical Analysis}

The blurring of images by the atmosphere and imperfections 
in telescope optics (seeing) degrades measurements of the 
surface brightnesses of galaxies. The seeing is characterized 
by the point spread function (PSF). The PSF gives the probability
 that a photon will hit the imaging device at a point different 
from where it would have hit in the absence of seeing. This
  can be determined observationally by studying the scattering of stellar 
  light. PSFs are well described by Gaussian functions, Gaussian functions with 
  exponential wings, linear superpositions of Gaussian functions, Moffat 
  functions, etc. (e.g. Moffat 1969; King 1971; Schweizer 1979). Among these 
  analytical approximations of PSFs, the most widely used is the single
   Gaussian. The main goal of the present paper is evaluate the effects of 
   seeing on the parameters of S\`ersic profiles in a completely analytical 
   form and to take into account the ellipticity of the surfaces brightness 
   distribution in this treatment. We develop our  analysis using a Gaussian PSF  and in
 Section 5 we shall compare this with a different PSF.

Assume a circular Gaussian function of dispersion $\sigma$ to model the point
 spread function:
\begin{equation}
{\rm PSF}(r)=\frac{1}{2 \pi \sigma^2}e^{-\frac{1}{2}(\frac{r}{\sigma})^2}.
\end{equation}
Consider a case where, in the absence of seeing, the surfaces brightness 
distribution $I({\bf {r}})$ of the galaxy is elliptically symmetric. This means 
that the isophotes of the object all have the same constant ellipticity 
$\epsilon$ ($\epsilon=1-b/a$, where $a$ and $b$ are the semi-major 
and semi-minor axes , respectively, of the isophote).

Elliptical coordinates ($\xi,\theta)$ are the most appropriate for our 
problem and are defined as:

\begin{eqnarray}
x&=&\xi \cos \theta 
\nonumber\\
y&=&\xi (1-\epsilon)\sin \theta
\end{eqnarray} 

In this coordinate system, 
the surface brightness distribution, $I({\bf {r}})$, of 
an elliptical source depends only on $\xi$: $I({\bf {r}})=I(\xi)$. 
The convolution equation that represents the effect of seeing on the 
surface brightness distribution is given by:

\begin{equation}
I_c(\xi,\theta) = (1-\epsilon) \int_0^\infty \xi^{'} d\xi^{'} \int_0^{2 \pi} 
d\theta^{'} {\rm PSF}(\xi^{'}, \theta^{'}, \xi, \theta) I(\xi^{'}),
\end{equation}
where PSF$(\xi^{'}, \theta^{'}, \xi, \theta)$ is the Gaussian PSF given by:

\begin{eqnarray}
\lefteqn{
 {\rm PSF}(\xi^{'}, \theta^{'}, \xi, \theta)=\frac{1}{2\pi \sigma^2} e^{-\frac{1}{2}\frac{\xi^2+\xi^{'2}}{\sigma^2}}
e^{\frac{\xi\xi{'}\cos(\theta+\theta^{'})}{\sigma^2}} {}}
\nonumber \\
& & {}e^{-\frac{(\epsilon^{2}-2\epsilon)(\xi^{'} \sin\theta^{'}-\xi \sin\theta)^2}{2\sigma^2}}.
\end{eqnarray}

\subsection{Analytical convolution of S\`ersic profiles}
In the particular case of S\`ersic profiles, the surface
 brightness distribution is given by: 

\begin{equation}
I(\xi)=I(0)e^{-(\frac{\xi}{r_0})^{\frac{1}{n}}},
\end{equation}
where $I(0)$ is the central intensity and $r_0$ is the scale length 
 of the profile. Over the major axis
of the object, $\theta=0$, the analytical solution of Eq. (3) for
 this type of profile can be written as:

\begin{eqnarray}
\lefteqn{I_c(\xi,0)=\frac{I(0)}{\pi^{1/2}}(1-\epsilon)e^{-\frac{1}{2}
(\frac{\xi}{\sigma})^{2}}\sum_{k=0}^{\infty}
\frac{(-)^k}{k!}\left(\frac{\sqrt{2}\sigma}{r_0}\right)^{\frac{k}{n}} 
\sum_{l=0}^{\infty}\frac{1}{l!}(2\epsilon-\epsilon^2)^l {}}
\nonumber\\
& {}\frac{\Gamma(l+\frac{1}{2})}{\Gamma(l+1)}\Gamma\left( l+1+\frac{k}{2n}\right) M\left( l+1+\frac{k}{2n},l+1,\frac{1}{2}
\left(\frac{\xi}{\sigma}\right)^{2}\right),
\end{eqnarray}
where $M(\mu,\nu,z)$ are the confluent hypergeometric functions
 (Abramowitz \& Stegun 1964, p. 504). This expression simplifies 
 if the object is circular:

\begin{eqnarray}
\lefteqn{I_c(\xi)=I(0)e^{-\frac{1}{2}
(\frac{\xi}{\sigma})^{2}}
\sum_{k=0}^{\infty}
\frac{(-)^k}{k!}\left(\frac{\sqrt{2}\sigma}{r_0}\right)^{\frac{k}{n}}
\Gamma\left(1+\frac{k}{2n}\right) {}}
\nonumber\\
& {}
M\left(1+\frac{k}{2n},1,\frac{1}{2}
\left(\frac{\xi}{\sigma}\right)^{2}\right).
\end{eqnarray}

In the limiting cases, $\xi \rightarrow \infty$ or $\sigma \rightarrow 0$, 
 the asymptotic expression of the confluent hypergeometric function 
 (Abramowitz \& Stegun 1964, p. 504) can be used to recover the original
  S\`ersic expression:

\begin{equation}
I_c(\xi,0)=I(0)e^{-(\frac{\xi}{r_0})^{\frac{1}{n}}}\left[1+O\left(\left(\frac{\xi^2}{2\sigma^2}\right)^{-1}\right)\right],
\end{equation}
where $(\xi^2/2\sigma^2)^{-1}$ quantifies the differences 
between the  S\`ersic profiles unaffected and affected by seeing in 
those asymptotic limits.

\subsection{The effect of seeing on the ellipticity of the isophotes}

In the absence of seeing, by construction, all isophotes of the 
S\`ersic profile have the same ellipticity, whereas the presence of seeing tends
to make them circular.
Using the isophote condition, $I(\xi,0)=I(\xi,\pi/2)$---the expression 
over the minor axis is written on appendix A---it is possible to derive 
an implicit equation that gives the 
variation of the ellipticity with the radial distance:

\begin{equation}
 \epsilon(\xi)=1-\left[-2\left(\frac{\sigma}{\xi}\right)^2\ln f(\epsilon, \sigma^2,
 n,r_0,\xi,\epsilon(\xi))\right]^{\frac{1}{2}},
\end{equation}
where f is given by:
\begin{eqnarray}
\lefteqn{ f(\epsilon, \sigma^2,
 n,r_0,\xi,\epsilon(\xi)))=\sigma^2e^{-\frac{1}{2}
(\frac{\xi}{\sigma})^{2}} {} }
\nonumber\\
 & {} \sum_{k=0}^{\infty}\frac{(-)^k}{k!}
(\frac{\sqrt{2}\sigma}{r_0})^{\frac{k}{n}} 
\sum_{l=0}^{\infty}\frac{1}{l!}(2\epsilon-\epsilon^2)^l
\frac{\Gamma(l+\frac{1}{2})}{\Gamma(l+1)}
\nonumber\\
&  {}\Gamma(l+1+\frac{k}{2n})M(l+1+\frac{k}{2n},l+1,\frac{1}{2}
(\frac{\xi}{\sigma})^{2}) 
\nonumber\\
&  {}
[ \sum_{n=0}^{\infty}\frac{1}{n!}\frac{(\epsilon^{2}-2\epsilon)^{n}}{(1-\epsilon)^{n}}
(\frac{1}{\xi(1-e(\xi))})^{n}\Gamma(n+\frac{1}{2})
\nonumber\\
 &  {}
\int_0^{\infty}e^{-(\frac{\xi^{'}}{r_0})^{\frac{1}{n}}}\xi^{'l+1}e^{-\frac{1}{2}
(\frac{\xi^{'}(1-\epsilon)}{\sigma})^{2}}I_n(\frac{\xi(1-\epsilon(\xi))\xi^{'}(1-\epsilon)}{\sigma^2}) ]^{-1} 
\end{eqnarray}
where $I_n(x)$ are the modified Bessel functions (Abramowitz
 \& Stegun 1964, p. 376).
\begin{figure}
  \vspace*{200pt}
  \caption{Effects of  seeing on the ellipticity. A model with
    $r_e/\sigma=2$ is shown.}
\end{figure}

Figure 1 shows the radial variation of the ellipticity 
due to the seeing. Note how the seeing affects  the central points, 
rounding the isophotes, whereas its effects are progressively less in 
the outer regions of the profile.

The equations that we have presented for the S\`ersic profiles can be
 immediately generalized to almost all the experimental profiles 
 (see the theorem in Appendix B).

\section{The effects of seeing on the S\`ersic profile parameters}

\subsection{The effect of seeing on the central intensity}

To study this effect we use  Eq. (6) and  apply the fact that
 at $\xi=0$ the confluent hypergeometric function satisfies 
 $M(\mu,\nu;0)=1$. This expression can then be written as:

\begin{eqnarray}
\lefteqn{I_c(0)=I(0)(1-\epsilon)\sum_{k=0}^{\infty}
\frac{(-)^k}{k!}\left(\frac{\sqrt{2}\sigma}{r_0}\right)^{\frac{k}{n}}
\Gamma\left( 1+\frac{k}{2n}\right) {}}
\nonumber\\
& {}{_2F_1}\left(\frac{1}{2},1+\frac{k}{2n};1;2\epsilon-\epsilon^2\right),
\end{eqnarray}

where $_2F_1(a,b;c,z)$ is the hypergeometric function
 (Abramowitz \& Stegun 1964, p. 556).

As shown in  Figure 2, the central intensity of the profile 
affected by seeing decreases monotonically when the seeing size increases. 
The central intensity of profiles with larger values of $n$ decreases more
 rapidly than for low $n$ as is expected because of the higher central 
 concentration of these profiles.

The central concentration of the object is also dependent on 
the ellipticity. In Figure 2 this relation is also shown. Larger 
ellipticities are more affected by  seeing.

\begin{figure}
  \vspace*{200pt}
  \caption{Effects of the seeing on the central intensity $I_c(0)$ for
  different values of $n$. 
  Three different ellipticities are shown, $\epsilon=0$ (full line),
   $\epsilon=0.25$ (dashed line) and $\epsilon=0.5$ (dotted dashed line).}
\end{figure}

\subsection{The effect of seeing on the effective radius}

The S\`ersic profile can be written in terms of the effective 
radius and the effective intensity:

\begin{equation}
I(r)=I_e10^{-b_n\left[\left(\frac{r}{r_e}\right)^{\frac{1}{n}}-1\right]}.
\end{equation}

The constant $b_n$ is chosen such that half the total luminosity predicted by
the law comes from  $r<r_e$. $b_n$ can be well approximated by the relation 
$b_n=0.868n-0.142$. $I_e$ is the intensity at the effective radius.

The relation between $I_e$, $r_e$ and $I(0)$, $r_0$ is given by:

\begin{eqnarray}
I(0)&=&I_e10^{b_n} \\
\noindent {\rm and}&& \nonumber \\
r_0&=&(b_n \ln 10)^{-n}r_e.
\label{cambios}
\end{eqnarray}

The seeing effect on effective radius can be obtained from 
the conservation of luminosity by the convolution:

\begin{equation}
L^c(r_e^c)=L(r_e),
\end{equation}
where $L(r_e)$ is the luminosity of the source inside $r_e$ and  $L^c(r_e^c)$
 is  the luminosity obtained from the object affected by seeing, measured inside 
 its effective radius.
For a circular object, $\epsilon=0$, we have

\begin{equation}
L^c(r_e^c)=2\pi\int_0^{r_e^c}rI_c(r)dr
\end{equation}
and $L(r_e)=I(0)\pi
r_e^2[n\Gamma(2n)]/[(b_n \ln 10)^{2n}]$ for a S\`ersic profile.
 Equation (15) can then be written  analyti\-cally for a circular system 
 as the implicit equation:

\begin{eqnarray}
\lefteqn{r_e^2\frac{n\Gamma(2n)}{(b_n \ln 10)^{2n}}= {r_e^c}^2 \sum_{k=0}^{\infty}\frac{(-)^k}{k!}
\left(\frac{\sqrt{2}\sigma}{r_0}\right)^{\frac{k}{n}}\Gamma\left(
1+\frac{k}{2n}\right) {}}
\nonumber\\
& {}\sum_{l=0}^{\infty}\frac{(-)^l}{l!}\frac{1}{l+1}\frac{\left(-\frac{k}{2n}\right)_l}{(1)_l}
\left(\frac{{r_e^c}^2}{2\sigma^2}\right)^l,
\end{eqnarray}
where $(\alpha)_l$ is the Pochhammer symbol: 
$(\alpha)_l\equiv\Gamma(\alpha+l)/\Gamma(\alpha)$. Figure 3 
shows that the effect of seeing is to increase the effective radius.
 This effect becomes more important as $n$ increases. The ellipticity
  effect is also shown; however, for $\epsilon\neq 0$ there is no easy 
  an analytical form, so the results with $\epsilon\neq 0$ that are 
  shown in Figure 3 were obtained numerically. Greater ellipticities
   imply greater effective radii, these differences are more important 
   for greater values of $n$. This result is as expected due to the
    diminution of the central intensity by the ellipticity effect.

\begin{figure}
  \vspace*{200pt}
  \caption{Effects of seeing on the effective radius, $r_e^c$. 
   Three different ellipticities for the source are shown:
    $\epsilon=0$ (full line), $\epsilon=0.25$ (dotted line) 
    and $\epsilon=0.5$ (dashed line).}
\end{figure}

It must be noted that our measurement of effective radius has 
been obtained over the semi-major axis. Some authors use as radial 
distance the magnitude $r^*=\sqrt{ab}$, in this case, the effective 
radius of the object affected by seeing is given by 
$r_e^{c*}=r_e^c\sqrt{1-\epsilon (r_e^c)}$, where $\epsilon(r_e^c)$ 
can be obtained from Equation (9).

\subsection{The effect of seeing on the parameter $n$}

To quantify the effect of seeing on the parameter $n$ we use the parameter:

\begin{equation}
\eta(\xi)\equiv\frac{1}{\xi}\frac{I(\xi)}
{\frac{dI(\xi)}{d\xi}} \ln \frac{I(\xi)}{I(0)}.
\end{equation}
This parameter is defined in such a way that $\eta(\xi)=n$ for 
all values of $\xi$ if $I(\xi)$ is a S\`ersic profile (Eq. 5). 
So $\eta(\xi)$ is equivalent, locally, to the parameter $n$ of
 the S\`ersic profile. Figure 4 summarizes the values of this
  parameter at $r_e^c$ 
  for different $n$ values. 
  It is easy to see that this parameter is the most affected by
   the seeing. Indeed, $\eta(0)=0.5$ for any profile affected by
    a Gaussian seeing. It should be noted that S\`ersic profiles 
    with $n=0.5$ are Gaussian profiles and then its convolution with a
     Gaussian PSF gives another Gaussian. The principal conclusion is
       that seeing effects always produce a surface brightness profile
        with a smaller value of $n$ than the actual value. 
Again, as $n$ increases,
	 the parameter is more affected. The effect of the ellipticity is to 
	 decrease the value of $n$ but the changes are not so important.

\begin{figure}
  \vspace*{200pt}
  \caption{Values of the parameter $\eta$ at $r_e^c$ as function of 
  the ratio $r_e^c/\sigma$ for different values of $n$ and ellipticities:
   $\epsilon=0$ (full line), $\epsilon=0.25$ (dashed line) and 
   $\epsilon=0.5$ (dotted-dashed line).}
\end{figure}

\section[]{A prescription for seeing corrections}

Here we present an easy prescription based on the use of the plots of Figures
2, 3, 4 and 5 (see below). This procedure permits 
the parameters of the S\`ersic profile (seeing-free quantities) to be
obtained using the
observational surface brightness profile. In summary, observers should:
\begin{description}
\item a) Determine the FWHM (Full Width at Half Maximum) of stars by fitting 
a Gaussian. $\sigma$ is related
to the FWHM by $\sigma=FWHM/\sqrt{8\ln 2}$.
\item b) Measure  $r_e^c$ along the semi-major axis solving the 
implicit equation
$L^c(r_e^c)=(1/2)L^c(\infty)$. This can be done (without any assumptions)
directly from the raw images.

\item c) Determine $\eta(r_e^c)$ numerically using the expression: 
\begin{equation}
\eta(r_e^c)=\frac{1}{r_e^c}\frac{I_c(r_e^c)}
{\frac{dI_c(\xi)}{d\xi}\arrowvert_{r_e^c}} \ln \frac{I_c(r_e^c)}{I_c(0)}.
\end{equation}

\item d) Evaluate the value of $n$ which corresponds to the 
point ($\eta(r_e^c)$,
$r_e^c/\sigma$) using Figure 4. Suppose, as a first approximation,
 that the value of $\epsilon$ corresponds to
the value of $\epsilon(r_e^c)$. Note that $\eta(\xi)$ is the
parameter less affected by the value of $\epsilon$, so the approximation is good.

\item e) Recalculate the value of $\epsilon$ more accurately using the value of $n$
obtained and Figure 5. Figure 5 shows the values of $\epsilon(r_e^c)$ for
different values of $n$.

\item f) Obtain the value of $r_e$ using Figure 3.

\item g) Obtain the value of $I(0)$ using Figure 2.
\end{description}
Observers wishing to be more precise can use the formulae instead the
figures.

\begin{figure}
  \vspace*{200pt}
  \caption{Values of the ellipticity of the isophotes at $r_e^c$ as function of 
  the ratio $r_e^c/\sigma$ for different values of $n$.}
\end{figure}

\section[]{Discussion}

The study presented here has assumed a Gaussian PSF for the seeing. 
 The real observed PSF is not exactly Gaussian. The theory of atmospheric 
 turbulence predicts the PSF to be the Fourier transform 
 of $e^{-(kb)^{5/3}}$ (Fried 1966; Woolf 1982), where $b$ is a 
 scaling parameter. Saglia et al. (1993) generalized this result,
  they assume a PSF  that is the Fourier transform of $e^{-(kb)^{\gamma}}$. 
  The Gaussian PSF is a particular case with $\gamma=2$. Their observational
   PSFs were in agreement with the theoretical PSF inferred by the turbulence
    theory. They  obtain  a Gaussian FWHM 4.67\% greater than the 
    FWHM of the turbulence PSF with $\gamma=5/3$. This systematic error will
     be transmitted into the parameters $I_{e}$, $r_{e}$ and $n$.  We have 
     computed these parameters, varying $\sigma$ in our PSF by 4.67\%,
      and have found that $I_{e}$ is underestimated by $7\%$, $r_{e}$ is 
      overestimated by $7\%$ and $n$ is underestimated by 4\% in a 
      systematic manner with respect to the initial values. Due to the 
      systematic character of these errors they can be easily taken 
      into account.

The parameter most affected by the seeing is $n$. This has
 important consequences because $n$ serves to classify the type of profile and
  is related to the central luminosity concentration. Graham et al. (1996) 
  and Jerjen, Bingelli \& Freeman (2000) found a correlation between
 $n$ and  galactic 
  type. Thus, dwarfs ellipticals show the smallest values of $n$ and cD
   galaxies have the largest values. Between these extremes
 are located the ellipticals
    and the bulges of spirals.   The parameter $\eta(\xi)$ defined in 
    Section 3.2 gives information locally about the value of $n$ over 
    the profile. Figure 4 shows $\eta(r_e^c)$ for different values of 
    the seeing. It can be observed that the seeing always produces 
    $\eta(r_e^c)<n$ for all the values of the ${r_e^c}/{\sigma}$ ratio.
     If  seeing is not taken into account, the value of $n$ that can be
      measured from the profiles is always smaller than the real one. 
      Usually,  fitting procedures avoid the central points in order
       to remove  seeing effects on the profiles. This is clearly not 
       sufficient to recover the real value of $n$ (see Figure 4).

One physical restriction to the values of $n$ is given by the luminosity 
density. For a homologous triaxial ellipsoid, the luminosity density 
(Stark 1977) associated with a S\`ersic profile is given by:

\begin{equation}
j(\zeta)=\frac{f^{\frac{1}{2}}}{\pi}\frac{K}{n}\frac{I(0)}{r_e^{1/n}}
\int_{\zeta}^{\infty}e^{-K\left(\frac{\xi}{r_e}\right)^{\frac{1}{n}}}\xi^{\frac{1}{n}-1}
\left(\xi^2-\zeta^2\right)^{-\frac{1}{2}}d\xi,
\end{equation}
where $K=b_n \ln 10$ and f$^{1/2}$ is a constant that 
depends on the 3D spatial orientation of the object. We have 
calculated the luminosity density for S\`ersic profiles with 
$n<1$ (see Figure 6). For $n<0.5$ the density has a depression 
in its central parts. This represents an unlikely physical situation.
 Nevertheless, the seeing effects prevent the measurement of $n<0.5$ 
 for objects with $n\geq0.5$ (see Figure 4).

\begin{figure}
  \vspace*{200pt}
  \caption{Normalized luminosity density $J(\zeta)=j(\zeta){r_e^{1/n}}/[f^{1/2}I(0)]$ vs. the radial coordinate 
  $\zeta/r_e$ for S\`ersic profiles with $n<1$: $n=0.2,0.3$ and 0.4 
  (dashed lines), $n=0.5$ (solid line) and $n=0.7, 0.8$ and 1.0 (dotted lines).}
\end{figure}

The effects of seeing on the parameters $I_{e}$, $r_{e}$ and $n$ depend 
on the intrinsic ellipticity ($\epsilon$) of the source. Thus, assuming that 
 the object has $\epsilon=0$, when it really is elliptical-symmetric, 
 results in the central intensity and $n$  begin underestimated whereas
  $r_{e}$ is overestimated. These effects are not negligible and they are 
  more important when $\epsilon$ increases.

\section{Conclusions}

We have developed an analytical study of the seeing effects on 
S\`ersic profiles.  The seeing PSF was modelled by a Gaussian function.
 In this analysis we have taken into account the intrinsic ellipticity 
 of the objects. Our main results are:

\begin{enumerate}
\item The convolved surface brightness profile along the major 
axis of the object can be expressed as a double series of confluent
 hypergeometric functions. This result is very general and can be applied 
 to nearly all the experimental surface brightness profiles.
\item We have obtained an implicit equation to evaluate the 
effect of seeing on the ellipticity of the isophotes. The rounding
 of the isophotes depends on $n, r_e, \sigma$ and $\epsilon$ in a 
 unique way.
\item The parameter most affected by the seeing effect is $n$. The 
observed S\`ersic profiles show smaller values of $n$, due to the 
seeing effect, than the real ones. Greater values of $n$ are the most 
affected. Also, for $n<0.5$, the luminosity density associated with a 
S\`ersic profile has a depression in its central parts. This represents
 an unlikely physical situation.
\item The seeing effects on the parameters of the S\`ersic profile 
depend on the intrinsic ellipticity of the object, therefore it is 
necessary to include it when  seeing effects are studied.

The results described here clearly show that seeing effects are
 important when one tries to measure accurate values of the parameters
  of a profile affected by the seeing. These results have to be taken into 
  account for sources with a low ${r_e}/{\sigma}$ relation as 
   expected for medium- and high-redshift objects.
\end{enumerate}
\section*{Acknowledgments}

We wish to thank J. A. Rubi\~no for valuable discussions and E.
 Simmoneau for useful comments. We are also indebted to V. Debattista 
 who kindly read versions of this manuscript. JALA was
supported by grant 20-56888.99 from the Schweizerischer Nationalfonds.

\appendix

\section[]{Analytical expression on minor axis}

Using Eq. (3) we can also get the equation for the minor axis,
 $\theta={\pi}/{2}$,

\begin{eqnarray}
\lefteqn{ I_c\left(\xi,\frac{\pi}{2}\right)=\frac{I(0)}{\pi^{1/2}}\frac{1}{(1-\epsilon)}e^{-\frac{1}{2}
\left[\frac{\xi(1-\epsilon)}{\sigma}\right]^{2}}\sum_{k=0}^{\infty}\frac{(-)^k}{k!}
\left[\frac{\sqrt{2}\sigma}{r_0(1-\epsilon)}\right]^{\frac{k}{n}} {}}
\nonumber\\
& {} \sum_{l=0}^{\infty}\frac{1}{l!}\left[\frac{\epsilon^2-2\epsilon}{(1-\epsilon)^{2}}\right]^l
\frac{\Gamma(l +\frac{1}{2})}{\Gamma(l+1)}\Gamma\left( l+1+\frac{k}{2n}\right) 
\nonumber\\
& {} M\left( l+1+\frac{k}{2n},l+1,\frac{1}{2}
\left[\frac{\xi(1-\epsilon)}{\sigma}\right]^{2}\right).
\end{eqnarray}
However, this expression is divergent for $\left|({\epsilon^2-2\epsilon})/{(1-\epsilon)^2}\right|>1$, so that 
for $\epsilon\geq$ 0.3 the use of the integral expression is required:
\begin{eqnarray}
\lefteqn{I_c\left(\xi,\frac{\pi}{2}\right) = \frac{1}{2 \pi \sigma^2}(1-\epsilon)e^{-\frac{1}{2}\frac{(1-\epsilon)^2\xi^2}{\sigma^2}} \int_0^\infty \xi^{'} d\xi^{'} 
I(\xi^{'}) e^{-\frac{1}{2}\frac{(1-\epsilon)^2\xi^{'2}}{\sigma^2}} {}}
\nonumber\\
& {}F_2(\xi,\xi^{'},\sigma,\epsilon),
\end{eqnarray}
where F$_2$($\xi,\xi^{'},\sigma,\epsilon$) is given by

\begin{eqnarray}
\lefteqn{F_2\left(\xi,\xi^{'},\sigma,\epsilon\right)=
2\pi^{\frac{1}{2}}\lbrace\pi^{\frac{1}{2}}I_0\left(\frac{\xi\xi^{'}}{\sigma^2}\right)+
\sum_{n=1}^{\infty}\frac{1}{n!}\frac{(\epsilon^{2}-2\epsilon)^{n}}{(1-\epsilon)^{2n}}
\left(\frac{\xi^{'}}{\xi}\right)^{n} {}}
\nonumber\\
& {}
\Gamma\left(n+\frac{1}{2}\right)I_n\left[\frac{\xi\xi^{'}}{\sigma^2}(1-\epsilon)^2\right]\rbrace .
\end{eqnarray}

In the limiting cases,  $\xi \rightarrow \infty$ or $\sigma \rightarrow 0$, for $\epsilon < 0.3$, the asymptotic expression of the confluent hypergeometric function (Abramowitz \& Stegun 1964) can be used again to recover the 
original S\`ersic  expression:

\begin{equation}
I_c\left(\xi,\frac{\pi}{2}\right)=I(0)e^{-\left(\frac{\xi}{r_0}\right)^{\frac{1}{n}}}\left[
1+O\left(\left[\frac{\xi^2}{2\sigma^2}(1-\epsilon)^2\right]^{-1}\right)\right] .
\end{equation}

\section[]{Theorem on Gaussian seeing}

Assume a surface brightness distribution with elliptical symmetry, 
I($\xi$). If this distribution can be written as the power series

\begin{equation}
I(\xi)=I(0) \sum _{k=0}^{\infty}\alpha (k)\xi^{\beta k},
\end{equation}
with a region of convergence equal to 0 $\leq$ $\xi$ $<$ $ \infty$, then the
 analytical solution of the convolution of $I(\xi)$ with a Gaussian PSF
  over the major axis is:

\begin{eqnarray}
\lefteqn{ I_c(\xi,0)=\frac{I(0)}{\pi^{1/2}}(1-\epsilon)e^{-\frac{1}{2}
\left(\frac{\xi}{\sigma}\right)^{2}}\sum_{k=0}^{\infty}
\alpha(k)\left(\sqrt{2}\sigma\right)^{\beta k} {}}
\nonumber\\
& {}\sum_{l=0}^{\infty}\frac{1}{l!}\left(2\epsilon-\epsilon^2\right)^l
\frac{\Gamma(l+\frac{1}{2})}{\Gamma(l+1)}\Gamma\left( l+1+\frac{\beta k}{2}\right) 
\nonumber\\
 & {}M\left( l+1+\frac{\beta k}{2},l+1,\frac{1}{2}
\left(\frac{\xi}{\sigma}\right)^{2}\right).
\end{eqnarray}

For $\epsilon$=0, the convolution can be written as

\begin{eqnarray}
\lefteqn{ I_c(\xi)=I(0)e^{-\frac{1}{2}
\left(\frac{\xi}{\sigma}\right)^{2}}\sum_{k=0}^{\infty}
\alpha (k)\left(\sqrt{2}\sigma\right)^{\beta k}
\Gamma\left( 1+\frac{\beta k}{2}\right) {}}
\nonumber\\
& {}
M\left( 1+\frac{\beta k}{2},1,\frac{1}{2}
\left(\frac{\xi}{\sigma}\right)^{2}\right).
\end{eqnarray}
For the S\`ersic profile,
 $\alpha(k)=[(-)^k/{k!}]({1}/{r_0})^{k/n}$, $\beta=1/n$.

Examples of profiles where the theorem is not applicable are the Hubble
 profile (Hubble 1930) and the Freeman bar profile (Freeman 1966). The
  first one does not admit a convergent power series over the entire interval 
  of definition, and the second one has a point of non-differentiability 
  that avoids a power series expansion. 

\bsp

\label{lastpage}

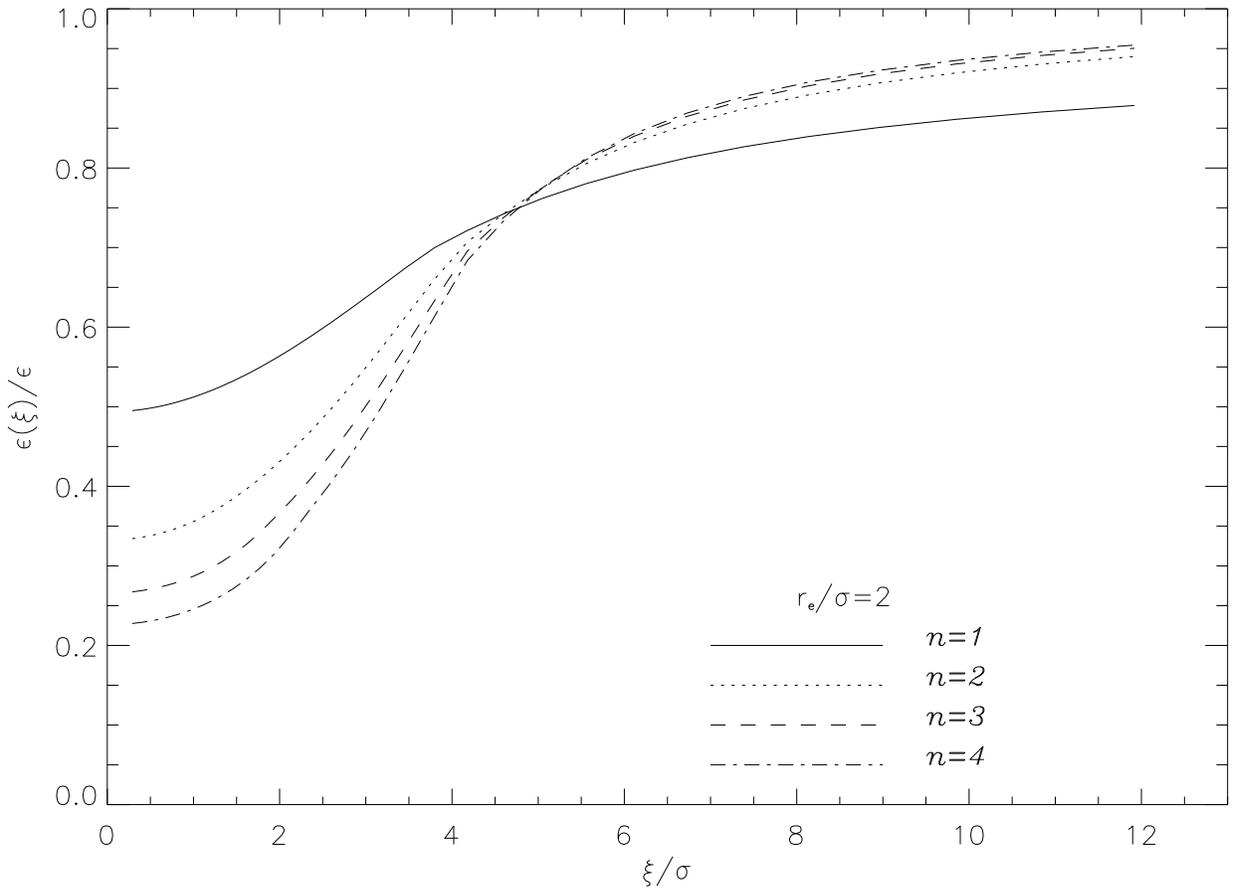
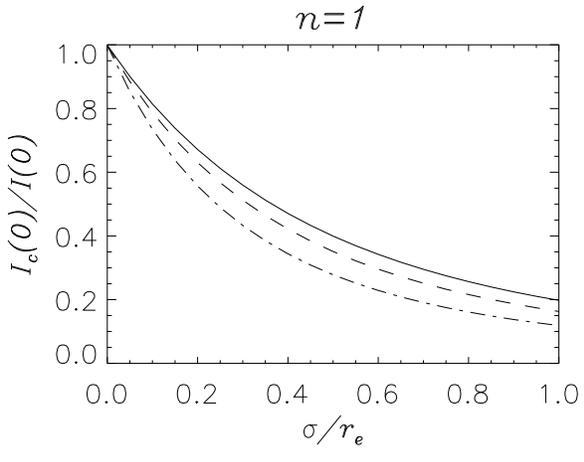
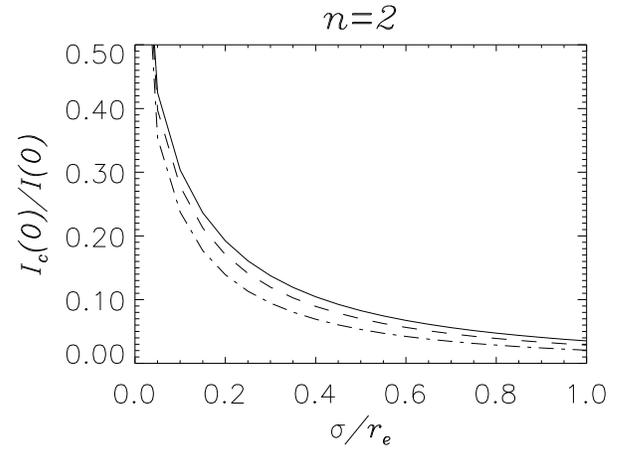
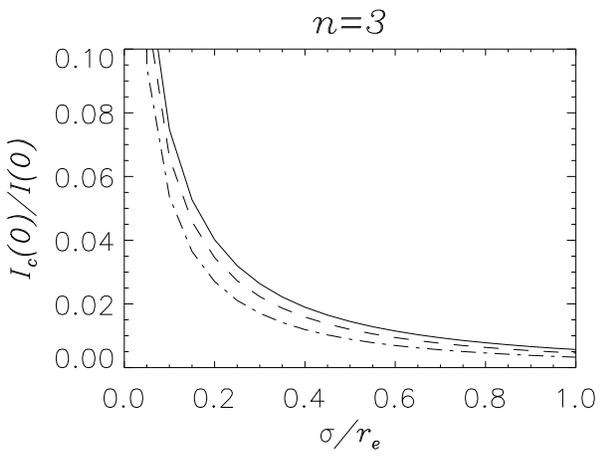
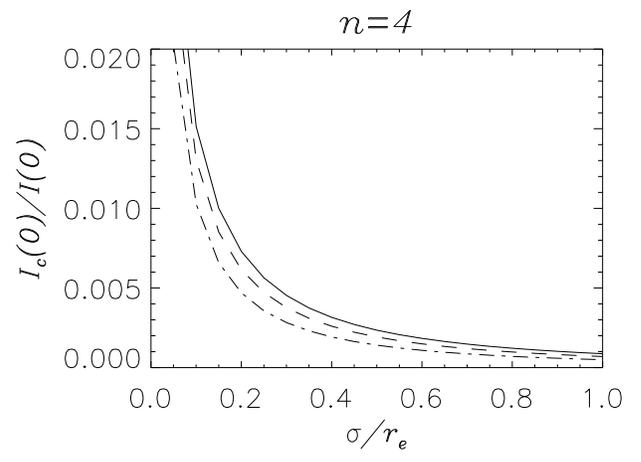
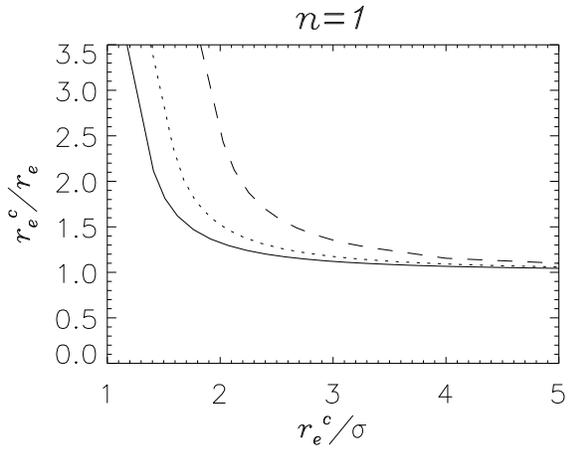
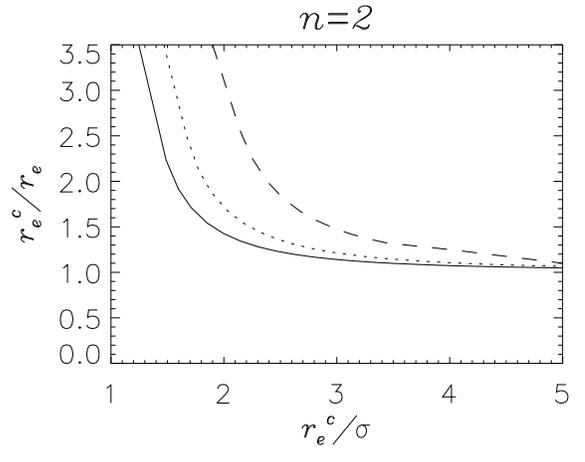
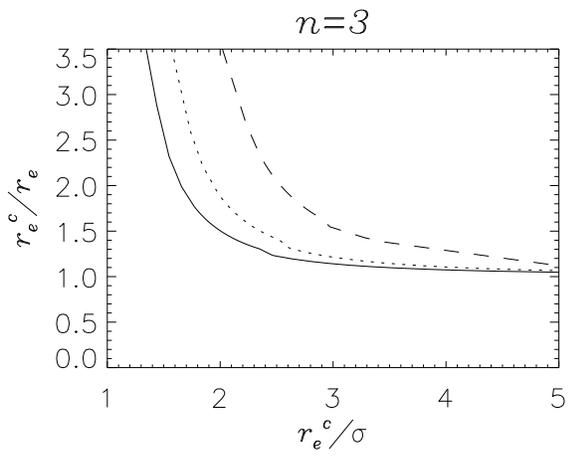
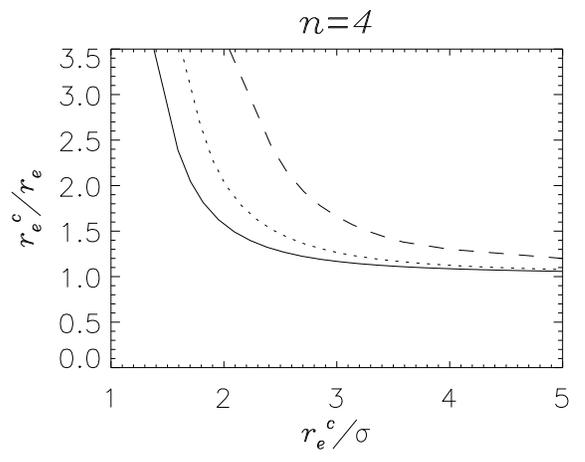
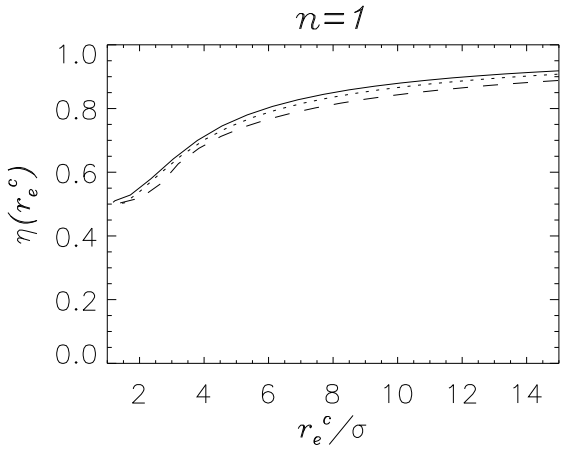
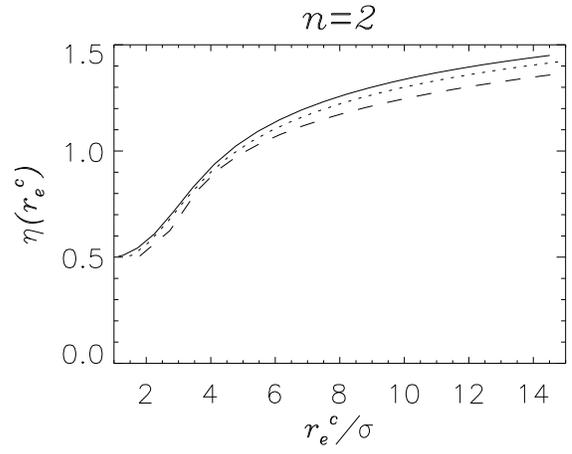
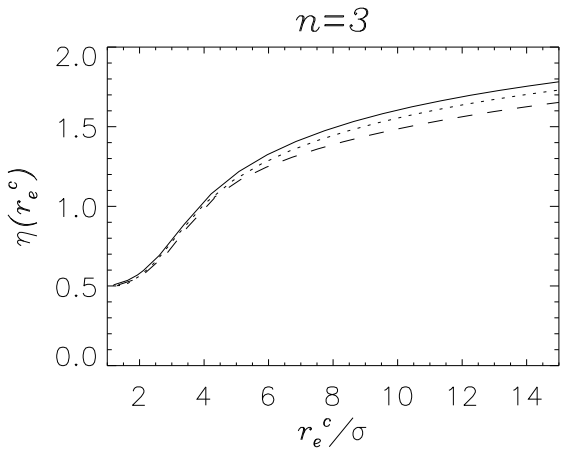
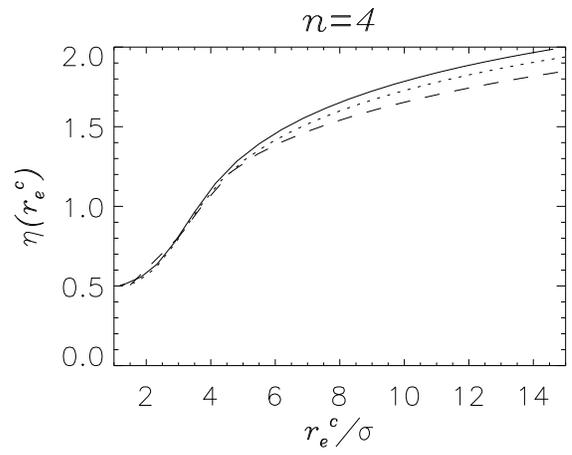
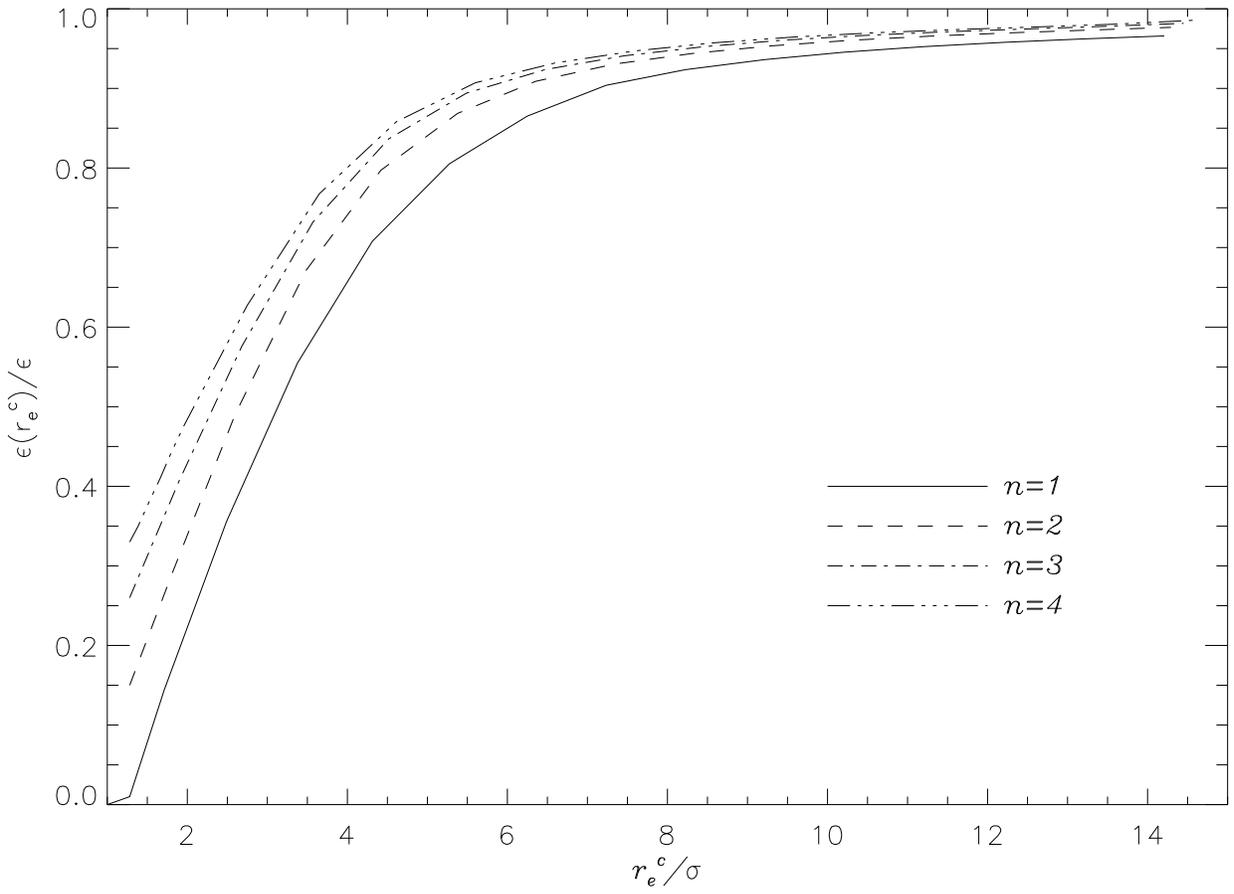
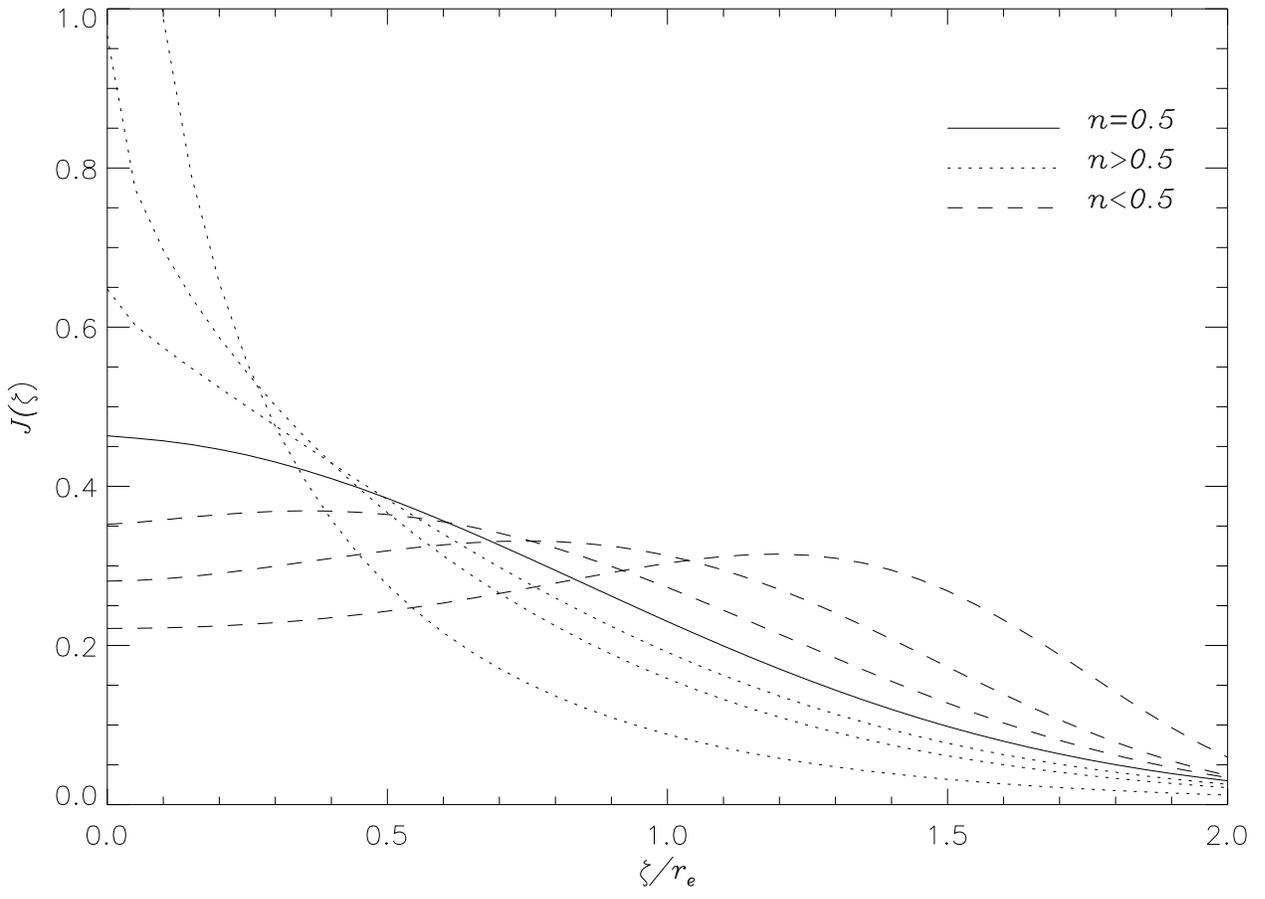
\end{document}